\documentclass[11pt,preprint]{revtex4}
\usepackage[dvips]{graphicx}
\usepackage{amsfonts}
\usepackage{slashed}
\usepackage{amsmath}

\def\E{\mathrm{E}}
\def\sech{\mathrm{sech}}

\begin{document}

\title{Induced Lorentz-violating terms at finite temperature}

\author{J. Leite}
\affiliation{Instituto de F\'\i sica, Universidade Federal de Alagoas, 57072-270, Macei\'o, Alagoas, Brazil}
\email{julioleite,tmariz@fis.ufal.br}

\author{T. Mariz}
\affiliation{Instituto de F\'\i sica, Universidade Federal de Alagoas, 57072-270, Macei\'o, Alagoas, Brazil}
\email{julioleite,tmariz@fis.ufal.br}

\date{\today}

\begin{abstract}
We study the radiatively induced Lorentz-violating terms at finite temperature, namely, the higher-derivative term and the Chern-Simons term. These terms are induced by integrating out the fermions coupled to the coefficient $g^{\kappa\mu\nu}$. The calculation of the resulting expressions is performed by using the derivative expansion and the Matsubara formalism. The Chern-Simons terms is nonzero only at finite temperature, whereas the higher-derivative term is finite at zero temperature, however, it goes to zero as the temperature grows to infinity. We also obtain a higher-derivative Chern-Simons term, nevertheless, it vanishes asymptotically.
\end{abstract}

\maketitle

\section{Introduction}

Studies of Lorentz violation have been recently performed in the literature through the standard-model extension \cite{Kos,Kos2,Kos3}, a effective field theory which contains terms formed by the contraction of Lorentz-violating operators of mass dimension $d=3$ and $d=4$ with coefficients of mass dimension $d=1$ and dimensionless, respectively. Operators of mass dimension $d\ge5$,  contracted with coefficients of mass dimension $d\le-1$, have also been considered \cite{Kos4}, however more analyses are still needed. In particular, studies at finite temperature were not taken into account yet. 

The aim of this work is to study the effect of a thermal bath on the first higher-derivative Lorentz-violating term, 
\begin{equation}\label{k5}
{\cal L}_k={\cal K}^{\mu\nu\rho\alpha\beta} A_\mu \partial_\rho\partial_\alpha\partial_\beta A_\nu,
\end{equation}
with
\begin{eqnarray}\label{K5}
{\cal K}^{\mu\nu\rho\alpha\beta} &=&k^{\mu\nu\alpha}g^{\beta\rho}+k^{\mu\nu\beta}g^{\alpha\rho}+k^{\nu\rho\alpha}g^{\beta\mu}+k^{\nu\rho\beta}g^{\alpha\mu}+k^{\rho\mu\alpha}g^{\beta\nu}+k^{\rho\mu\beta}g^{\alpha\nu}.
\end{eqnarray} 
The coefficient $k^{\mu\nu\alpha}$ has antisymmetry in the first two indices, so that the coefficient ${\cal K}^{\mu\nu\rho\alpha\beta}$ has antisymmetry in the first three indices and symmetry in the last two; both coefficients have mass dimension $d=-1$. Indeed, this term is radiatively generated by the derivative term $\bar\psi {i\over2}g^{\kappa\lambda\mu}\sigma_{\kappa\lambda}(\partial_\mu+iA_\mu)\psi$ of the Lorentz-violating QED, with $k^{\lambda\mu\nu}\propto m^{-1}g^{\lambda\mu\nu}$ \cite{Mar}. However, it was observed in \cite{Mar} that only the component $g^{0ij}$ induces a term in which the resulting theory appears to be unitarity.

Similar terms, involving operators of mass dimension $d=5$, have also been studied recently in the literature \cite{Sud,Mye,Bol,Rey}, however, as we mention above, studies at finite temperature never were performed. Because of the apparent analogy between the terms, we believe that they will exhibit similar behavior at finite temperature.

In addition to the study of the term (\ref{k5}), we will take into account the possibility of the induction of a Chern-Simons term, at finite temperature, given by
\begin{equation}\label{k3}
{\cal L}_\kappa={\cal K}^{\mu\nu\rho} A_\mu \partial_\rho A_\nu,
\end{equation}
where
\begin{equation}\label{K3}
{\cal K}^{\mu\nu\rho} =\kappa^{\mu\nu\rho}+\kappa^{\nu\rho\mu}+\kappa^{\rho\mu\nu}.
\end{equation} 
Here we have $\kappa^{\lambda\mu\nu}\propto m\,g^{\lambda\mu\nu}$, in order that ${\cal K}^{\mu\nu\rho}$ is totally antisymmetric. Note that ${\cal K}^{\mu\nu\rho}$ as well as $\kappa^{\lambda\mu\nu}$ has dimension of mass $d=1$, so that we can rewrite ${\cal K}^{\mu\nu\rho}=\epsilon^{\mu\nu\rho\sigma}(k_{AF})_\sigma$, or $(k_{AF})_\alpha=\frac{1}{3!}\epsilon_{\alpha\mu\nu\rho}{\cal K}^{\mu\nu\rho}$, where $(k_{AF})_\sigma$ is the usual coefficient for the Lorentz-violating Chern-Simons term. This term has been extensively studied \cite{Car,Col,Chu,Jac,Per,Bon,Ada,And,Alf} (see also references therein), even at finite temperature \cite{Cer,Ebe,Mar2,Gom,Cas}. 


\section{Chern-Simons term}\label{cs}

In this section, we are interested in studying whether the Chern-Simons term (\ref{k3}) is induced at finite temperature \footnote{For the three-dimensional Chern-Simons term, this issue has already been extensively studied in the literature \cite{Bab,Des,Fos}.} by integrating out the fermions coupled to the coefficient $g^{\kappa\mu\nu}$. As pointed out in \cite{Mar}, at zero temperature this term is not induced, only the higher-derivative term (\ref{k5}) is obtained in the low-energy limit.  

To study this issue, let us consider the fermionic sector of the Lorentz-violating QED described by the Lagrangian density
\begin{equation}\label{Lf}
{\cal L}_f = \bar\psi\left(i\slashed{\partial}+\frac i2 g^{\kappa\lambda\mu}\sigma_{\kappa\lambda}\partial_\mu-m-\slashed{A}-\frac12 g^{\kappa\lambda\mu}\sigma_{\kappa\lambda}A_\mu\right)\psi,
\end{equation}
where $\sigma_{\kappa\lambda}=\frac i2 [\gamma_\kappa,\gamma_\lambda]$. The corresponding Feynman rules and  the relevant one-loop contributions can be found in Ref.~\cite{Mar}. The resulting expressions are given by
\begin{subequations}\label{Pi}
\begin{eqnarray}
\label{Pia}i\Pi^{\mu\nu}_{(a)} &=& - \int\frac{d^4p}{(2\pi)^4} \mathrm{tr}\, (-i)\gamma^\mu iS(p)\frac i2 g^{\kappa\lambda\rho}\sigma_{\kappa\lambda}p_\rho iS(p)(-i)\gamma^\nu iS(p-k), \\
\label{Pib}i\Pi^{\mu\nu}_{(b)} &=& - \int\frac{d^4p}{(2\pi)^4} \mathrm{tr}\, (-i)\gamma^\mu iS(p)(-i)\gamma^\nu iS(p-k)\frac i2 g^{\kappa\lambda\rho}\sigma_{\kappa\lambda}(p_\rho-k_\rho) iS(p-k), \\
\label{Pic}i\Pi^{\mu\nu}_{(c)} &=& - \int\frac{d^4p}{(2\pi)^4} \mathrm{tr}\, \left(-\frac i2\right) g^{\kappa\lambda\mu}\sigma_{\kappa\lambda} iS(p)(-i)\gamma^\nu iS(p-k), \\
\label{Pid}i\Pi^{\mu\nu}_{(d)} &=& - \int\frac{d^4p}{(2\pi)^4} \mathrm{tr}\,  (-i)\gamma^\mu iS(p)\left(-\frac i2\right) g^{\kappa\lambda\nu}\sigma_{\kappa\lambda} iS(p-k),
\end{eqnarray}
\end{subequations}
with $S(p)=(\slashed{p}-m)^{-1}$ and $\mathrm{tr}$ means the trace over the Dirac matrices.  

The calculation of the integrals of the expressions (\ref{Pi}), at zero temperature, was performed in \cite{Mar} using Feynman parameterization. However, at finite temperature, in order to avoid employing this parameterization, it is more suitable to use the expansion of the fermionic propagator, 
\begin{equation}
S(p-k)=S(p)+S(p)\slashed{k}S(p)+\cdots, 
\end{equation}
and thus restrict ourselves to terms linear in $k$, which are sufficient to generate the Chern-Simons term. Then, the expressions (\ref{Pi}) became
\begin{subequations}\label{PI}
\begin{eqnarray}
\label{Pi1}\Pi^{\mu\nu}_{(1)} &=& -\frac i2 \int\frac{d^4p}{(2\pi)^4} \mathrm{tr}\, \gamma^\mu S(p)g^{\kappa\lambda\rho}\sigma_{\kappa\lambda}p_\rho S(p)\gamma^\nu S(p)\slashed{k}S(p), \\
\label{Pi2}\Pi^{\mu\nu}_{(2)} &=& -\frac i2 \int\frac{d^4p}{(2\pi)^4} \mathrm{tr}\, \gamma^\mu S(p)\gamma^\nu S(p)\slashed{k}S(p)g^{\kappa\lambda\rho}\sigma_{\kappa\lambda}\,p_\rho S(p), \\
\label{Pi3}\Pi^{\mu\nu}_{(3)} &=& \frac i2 \int\frac{d^4p}{(2\pi)^4} \mathrm{tr}\, \gamma^\mu S(p)\gamma^\nu S(p)g^{\kappa\lambda\rho}\sigma_{\kappa\lambda}k_\rho S(p), \\
\label{Pi3}\Pi^{\mu\nu}_{(4)} &=& -\frac i2 \int\frac{d^4p}{(2\pi)^4} \mathrm{tr}\, \gamma^\mu S(p)\gamma^\nu S(p)g^{\kappa\lambda\rho}\sigma_{\kappa\lambda}\,p_\rho S(p)\slashed{k}S(p), \\
\label{Pi5}\Pi^{\mu\nu}_{(5)} &=& \frac i2 \int\frac{d^4p}{(2\pi)^4} \mathrm{tr}\, g^{\kappa\lambda\mu}\sigma_{\kappa\lambda} S(p)\gamma^\nu S(p)\slashed{k}S(p), \\
\label{Pi6}\Pi^{\mu\nu}_{(6)} &=& \frac i2 \int\frac{d^4p}{(2\pi)^4} \mathrm{tr}\,  \gamma^\mu S(p) g^{\kappa\lambda\nu}\sigma_{\kappa\lambda} S(p)\slashed{k}S(p),
\end{eqnarray}
\end{subequations}
with $\Pi^{\mu\nu}_{(a,c,d)}\to\Pi^{\mu\nu}_{(1,5,6)}$, respectively, and $\Pi^{\mu\nu}_{(b)}\to\Pi^{\mu\nu}_{(2)}+\Pi^{\mu\nu}_{(3)}+\Pi^{\mu\nu}_{(4)}$.

Before assuming that the system is in thermal equilibrium with a temperature $T=\beta^{-1}$, let us first calculate the trace over Dirac matrices, so that we obtain
\begin{eqnarray}\label{Pikappa}
\Pi_\kappa^{\mu\nu} &=&  8m\,{\cal G}^{\mu\nu\rho\alpha\beta}k_\rho \int \frac{d^4p}{(2\pi)^4}\frac{p_\alpha p_\beta}{(p^2-m^2)^3}-4m\,{\cal G}^{\mu\nu\rho}k_\rho\int \frac{d^4p}{(2\pi)^4}\frac{1}{(p^2-m^2)^2},
\end{eqnarray}
where $\Pi_\kappa^{\mu\nu}=\Pi^{\mu\nu}_{(1)}+\Pi^{\mu\nu}_{(2)}+\Pi^{\mu\nu}_{(3)}+\Pi^{\mu\nu}_{(4)}+\Pi^{\mu\nu}_{(5)}+\Pi^{\mu\nu}_{(6)}$, 
\begin{eqnarray}\label{G5}
{\cal G}^{\mu\nu\rho\alpha\beta} &=& g^{\mu\nu\alpha}g^{\beta\rho}+g^{\mu\nu\beta}g^{\alpha\rho}+g^{\nu\rho\alpha}g^{\beta\mu}+g^{\nu\rho\beta}g^{\alpha\mu}+g^{\rho\mu\alpha}g^{\beta\nu}+g^{\rho\mu\beta}g^{\alpha\nu},
\end{eqnarray} 
and
\begin{equation}\label{G3}
{\cal G}^{\mu\nu\rho} =g^{\mu\nu\rho}+g^{\nu\rho\mu}+g^{\rho\mu\nu}.
\end{equation} 
In order to arrive at the expression (\ref{Pikappa}), we needed to simplify the scalar propagator $G(p)=(p^2-m^2)^{-1}$ in all expression by making the substitutions $p^2=G^{-1}(p)+m^2$ and $p^4=G^{-2}(p)+2G^{-1}(p)m^2+m^4$.

Now, it is interesting to change the Minkowski space to Euclidean space by performing the Wick rotation $p_0\to ip_0$ ($g^{\mu\nu}\to-\delta^{\mu\nu}$), i.e., $d^4p\to id^4p_\E=idp_0dp_1dp_2dp_3$, $p^2\to -\delta^{\mu\nu}p_\E^\mu p_\E^\nu = -p_\E^2$, $p\cdot k\to -p_\E\cdot k_\E$, $g^{\mu\nu\alpha}g_{\alpha\beta}p^\beta \to -g_\E^{\mu\nu\alpha}\delta^{\alpha\beta}p_\E^\beta = -g_\E^{\mu\nu\alpha}p_\E^\alpha$, and $g^{\mu\nu\rho}k_\rho \to -g_\E^{\mu\nu\rho}k_\E^\rho$. The resulting expression takes the form
\begin{eqnarray}
\Pi_\kappa^{\mu\nu} &=&  -8im\,{\cal G}_\E^{\mu\nu\rho\alpha\beta}k_\E^\rho \int \frac{d^4p_\E}{(2\pi)^4}\frac{p_\E^\alpha p_E^\beta}{(p_\E^2+m^2)^3}+4im\,{\cal G}_\E^{\mu\nu\rho}k_\E^\rho\int \frac{d^4p_\E}{(2\pi)^4}\frac{1}{(p_\E^2+m^2)^2},
\end{eqnarray}
with, obviously, 
\begin{eqnarray}\label{G5}
{\cal G}_\E^{\mu\nu\rho\alpha\beta} &=& g_\E^{\mu\nu\alpha}\delta^{\beta\rho}+g_\E^{\mu\nu\beta}\delta^{\alpha\rho}+g_\E^{\nu\rho\alpha}\delta^{\beta\mu}+g_\E^{\nu\rho\beta}\delta^{\alpha\mu}+g_\E^{\rho\mu\alpha}\delta^{\beta\nu}+g_\E^{\rho\mu\beta}\delta^{\alpha\nu}
\end{eqnarray} 
and
\begin{equation}\label{G3}
{\cal G}_\E^{\mu\nu\rho} =g_\E^{\mu\nu\rho}+g_\E^{\nu\rho\mu}+g_\E^{\rho\mu\nu}.
\end{equation} 

In order to calculate the space integral of the above equation we first decompose $ p_\E^\alpha=(p_0,p_i)$, as follows
\begin{equation}\label{decom}
 p_\E^\alpha=\hat p^\alpha + p_0\delta^{\alpha0},
\end{equation}
where, as a consequence, $\hat p^\alpha=(0,p_i)$. We also promote the $3$-dimensional space integral to $D$ dimensions, as well as introduce an arbitrary parameter $\mu$ to keep the mass dimension unchanged, and finally, due to the symmetry of the integral under spacial rotations, we replace
\begin{equation}\label{rep}
\hat p^\alpha \hat p^\beta \to \frac{\hat p^2}{D}(\delta^{\alpha\beta}-\delta^{\alpha0}\delta^{\beta0}).
\end{equation}
These procedures lead to the result
\begin{eqnarray}\label{Pikappa2}
\Pi_\kappa^{\mu\nu} &=&  i2^{1-D}m\,\pi^{-D/2}(\mu^2)^{\frac32-\frac D2}\,{\cal G}_\E^{\mu\nu\rho00}k_\E^\rho \Gamma\left(2-\frac D2\right) \nonumber\\
&&\times \int \frac{dp_0}{2\pi}\left[\frac{(D-3)}{(p_0^2+m^2)^{2-\frac D2}}-\frac{(D-4)m^2}{(p_0^2+m^2)^{3-\frac D2}}\right].
\end{eqnarray}
If we then calculate the $p_0$ integral, $\Pi_\kappa^{\mu\nu}=0$, as expected.

We are now in position to employ the Matsubara formalism, which consist in taking $p_0=(n+1/2)2\pi/\beta$ and changing $(1/2\pi)\int dp_0\to 1/\beta \sum_n$. To perform the summations, we cannot readily take the limit $D\to3$ in the Eq.~(\ref{Pikappa2}), because the sum of the first term exhibits singularities. Thus, in order to isolate these singularities, let us use an explicit representation for the sum over the Matsubara frequencies \cite{For}, given by
\begin{equation}\label{sum}
\sum_n\bigl[(n+b)^2+a^2\bigl]^{-\lambda} = \frac{\sqrt{\pi}\Gamma(\lambda-1/2)}{\Gamma(\lambda)(a^2)^{\lambda-1/2}}+4\sin(\pi\lambda)f_\lambda(a,b)
\end{equation}
where
\begin{equation}\label{f}
f_\lambda(a,b) = \int^{\infty}_{|a|}\frac{dz}{(z^2-a^2)^{\lambda}}Re\Biggl(\frac{1}{e^{2\pi(z+ib)}-1}\Biggl),
\end{equation}
which is valid for $Re\,\lambda<1$, aside from the poles at $\lambda=1/2,-1/2,-3/2,\cdots$. Then, using the above expression,  for the first sum of Eq.~(\ref{Pikappa2}), we have $\lambda\to1/2$ (when $D\to3$); however, for the second one, $\lambda\to3/2$, which clearly is out of range of validity. Therefore, we must use the recurrence relation
\begin{eqnarray}\label{rc}
f_{\lambda}(a,b) &=& -\frac1{2a^2}\frac{2\lambda-3}{\lambda-1}f_{\lambda-1}(a,b) - \frac1{4a^2}\frac1{(\lambda-2)(\lambda-1)}\frac{\partial^2}{\partial b^2}f_{\lambda-2}(a,b)
\end{eqnarray}
in the second sum of Eq.~(\ref{Pikappa2}), so that we have now $\lambda-1\to1/2$ and $\lambda-2\to-1/2$, as required for using Eq.~(\ref{f}). 

\begin{figure}[h]
\includegraphics[scale=1.0]{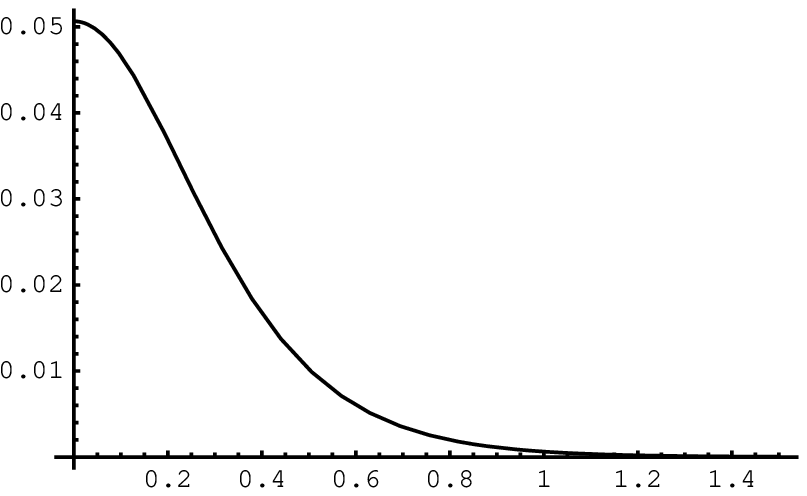}
\caption{Plot of the function $F(\xi)$} \label{F2}
\end{figure}

Finally, we obtain the following expression:
\begin{equation}
\Pi_\kappa^{\mu\nu} = \frac{im}2\,{\cal G}^{\mu\nu\rho00}k^\rho\,F(\xi),
\end{equation}
with ${\cal G}^{\mu\nu\rho00}=2g^{\mu\nu0}g^{0\rho}+2g^{\nu\rho0}g^{0\mu}+2g^{\rho\mu0}g^{0\nu}$ and
\begin{equation}\label{F}
F(\xi) = \int_{|\xi|}^\infty dz (z^2-\xi^2)^{1/2}\,\sech^2(\pi\xi)\tanh(\pi\xi),
\end{equation}
where $\xi=\beta m/2\pi$. Thus, re-installing the electron charge into the covariant derivative, $D_\mu=\partial_\mu+ieA_\mu$, we have 
\begin{equation}\label{CS}
{\cal L}_\kappa=\frac {e^2m}2F(\xi)\,{\cal G}^{\mu\nu\rho00} A_\mu \partial_\rho A_\nu,
\end{equation}
in which the plot of the function $F(\xi)$ is presented in Fig.~\ref{F2}.

The result (\ref{CS}) resembles that obtained previously in \cite{Cer}, when the fermions are integrated with the term $\bar\psi\slashed{b}\gamma_5\psi$ included. In fact, we can rewrite the Eq.~(\ref{CS}) as
\begin{equation}
{\cal L}_\kappa=\frac m2F(\xi)\,b_i\epsilon^{i\mu\nu\rho} A_\mu \partial_\rho A_\nu,
\end{equation}
where $b_i=\frac1{3!}\epsilon_{i\alpha\beta\gamma}{\cal G}^{\alpha\beta\gamma}$, with $g^{ij0}\ne0$ and $g^{0ij}=0$. This expression is consistent with the fact that, under a certain fermion field redefinition, the totally antisymmetric component of $g^{\mu\nu\rho}$ is completely absorbed into $b^\mu$ \cite{Kos2,McD}. Therefore, although the corresponding operators of the coefficients $g^{\mu\nu\rho}$ and $b^\mu$ are different, we also expect a radiative contribution coming from $g^{\mu\nu\rho}$ for the Lorentz-violating Chern-Simons term.

\section{Higher-derivative term}\label{hd}

Let us now study the effect of a thermal bath on the structure of the higher-derivative term (\ref{k5}). For this, we must expand the fermionic propagator $S(p-k)$ up to third order in $k$, in the expressions (\ref{Pi}), and single out the contributions  
\begin{subequations}\label{PI2}
\begin{eqnarray}
\label{Pii}\Pi^{\mu\nu}_{(i)} &=& -\frac i2 \int\frac{d^4p}{(2\pi)^4} \mathrm{tr}\, \gamma^\mu S(p)g^{\kappa\lambda\rho}\sigma_{\kappa\lambda}p_\rho S(p)\gamma^\nu S(p)\slashed{k}S(p)\slashed{k}S(p)\slashed{k}S(p), \hspace{1cm}\\
\label{Piii}\Pi^{\mu\nu}_{(ii)} &=& -\frac i2 \int\frac{d^4p}{(2\pi)^4} \mathrm{tr}\, \gamma^\mu S(p)\gamma^\nu S(p)\slashed{k}S(p)\slashed{k}S(p)\slashed{k}S(p)g^{\kappa\lambda\rho}\sigma_{\kappa\lambda}\,p_\rho S(p), \\
\label{Piiii}\Pi^{\mu\nu}_{(iii)} &=& -\frac i2 \int\frac{d^4p}{(2\pi)^4} \mathrm{tr}\, \gamma^\mu S(p)\gamma^\nu S(p)\slashed{k}S(p)\slashed{k}S(p)g^{\kappa\lambda\rho}\sigma_{\kappa\lambda}\,p_\rho S(p)\slashed{k}S(p), \\
\label{Piiv}\Pi^{\mu\nu}_{(iv)} &=& -\frac i2 \int\frac{d^4p}{(2\pi)^4} \mathrm{tr}\, \gamma^\mu S(p)\gamma^\nu S(p)\slashed{k}S(p)g^{\kappa\lambda\rho}\sigma_{\kappa\lambda}\,p_\rho S(p)\slashed{k}S(p)\slashed{k}S(p), \\
\label{Piv}\Pi^{\mu\nu}_{(v)} &=& -\frac i2 \int\frac{d^4p}{(2\pi)^4} \mathrm{tr}\, \gamma^\mu S(p)\gamma^\nu S(p)g^{\kappa\lambda\rho}\sigma_{\kappa\lambda}\,p_\rho S(p)\slashed{k}S(p)\slashed{k}S(p)\slashed{k}S(p),
\end{eqnarray}
\end{subequations}
and
\begin{subequations}\label{PI3}
\begin{eqnarray}
\label{Pivi}\Pi^{\mu\nu}_{(vi)} &=& \frac i2 \int\frac{d^4p}{(2\pi)^4} \mathrm{tr}\, \gamma^\mu S(p)\gamma^\nu S(p)\slashed{k}S(p)\slashed{k}S(p)g^{\kappa\lambda\rho}\sigma_{\kappa\lambda}\,k_\rho S(p), \\
\label{Pivii}\Pi^{\mu\nu}_{(vii)} &=& \frac i2 \int\frac{d^4p}{(2\pi)^4} \mathrm{tr}\, \gamma^\mu S(p)\gamma^\nu S(p)\slashed{k}S(p)g^{\kappa\lambda\rho}\sigma_{\kappa\lambda}k_\rho S(p)\slashed{k}S(p), \\
\label{Piviii}\Pi^{\mu\nu}_{(viii)} &=& \frac i2 \int\frac{d^4p}{(2\pi)^4} \mathrm{tr}\, \gamma^\mu S(p)\gamma^\nu S(p)g^{\kappa\lambda\rho}\sigma_{\kappa\lambda}\,k_\rho S(p)t\slashed{k}S(p)\slashed{k}S(p), \\
\label{Piix}\Pi^{\mu\nu}_{(ix)} &=& \frac i2 \int\frac{d^4p}{(2\pi)^4} \mathrm{tr}\, g^{\kappa\lambda\mu}\sigma_{\kappa\lambda} S(p)\gamma^\nu S(p)\slashed{k}S(p)\slashed{k}S(p)\slashed{k}S(p), \\
\label{Pix}\Pi^{\mu\nu}_{(x)} &=& \frac i2 \int\frac{d^4p}{(2\pi)^4} \mathrm{tr}\,  \gamma^\mu S(p) g^{\kappa\lambda\nu}\sigma_{\kappa\lambda} S(p)\slashed{k}S(p)\slashed{k} S(p)\slashed{k}S(p),
\end{eqnarray}
\end{subequations}
where $\Pi^{\mu\nu}_{(a,c,d)}\to\Pi^{\mu\nu}_{(i,ix,x)}$, respectively, and $\Pi^{\mu\nu}_{(b)}\to\Pi^{\mu\nu}_{(ii)}+\Pi^{\mu\nu}_{(iii)}+\Pi^{\mu\nu}_{(iv)}+\Pi^{\mu\nu}_{(v)}+\Pi^{\mu\nu}_{(vi)}+\Pi^{\mu\nu}_{(vii)}+\Pi^{\mu\nu}_{(viii)}$. Following the same above steps, namely, by calculating the trace and simplifying the scalar propagator in all expressions (\ref{PI2}) and (\ref{PI3}), with also the help of the substitution $p^6=G^{-3}(p)+3G^{-2}(p)m^2+3G^{-1}(p)m^4+m^6$, we obtain 
\begin{eqnarray}
\Pi_k^{\mu\nu} &=&  64m\,{\cal G}^{\mu\nu\rho\alpha\beta}k^\gamma k^\delta k_\rho \int \frac{d^4p}{(2\pi)^4}\frac{p_\alpha p_\beta p_\gamma p_\delta}{(p^2-m^2)^5}-32m\,{\cal G}^{\mu\nu\rho\alpha\beta}k^\gamma k_\beta k_\rho \int \frac{d^4p}{(2\pi)^4}\frac{p_\alpha p_\gamma}{(p^2-m^2)^4} \nonumber\\ 
&&-(12m\,{\cal G}^{\mu\nu\rho\alpha\beta}k^2 k_\rho+16m\,{\cal G}^{\mu\nu\rho}k^\alpha k^\beta k_\rho)\int \frac{d^4p}{(2\pi)^4}\frac{p_\alpha p_\beta}{(p^2-m^2)^4} \nonumber\\
&&+(4m\,{\cal G}^{\mu\nu\rho\alpha\beta}k_\alpha k_\beta k_\rho+4m\,{\cal G}^{\mu\nu\rho}k^2 k_\rho)\int \frac{d^4p}{(2\pi)^4}\frac{1}{(p^2-m^2)^3},
\end{eqnarray}
or changing from the Minkowski space to Euclidean space, we find
\begin{eqnarray}
\Pi_k^{\mu\nu} &=&  -64im\,{\cal G}_\E^{\mu\nu\rho\alpha\beta}k_\E^\gamma k_\E^\delta k_\E^\rho \int \frac{d^4p_\E}{(2\pi)^4}\frac{p_\E^\alpha p_\E^\beta p_\E^\gamma p_\E^\delta}{(p_\E^2+m^2)^5}+32im\,{\cal G}_\E^{\mu\nu\rho\alpha\beta}k_\E^\gamma k_\E^\beta k_\E^\rho \int \frac{d^4p_\E}{(2\pi)^4}\frac{p_\E^\alpha p_\E^\gamma}{(p_\E^2+m^2)^4} \nonumber\\ 
&&+(12im\,{\cal G}_\E^{\mu\nu\rho\alpha\beta}k_\E^2 k_\E^\rho+16im\,{\cal G}_\E^{\mu\nu\rho}k_\E^\alpha k_\E^\beta k_\E^\rho)\int \frac{d^4p_\E}{(2\pi)^4}\frac{p_\E^\alpha p_\E^\beta}{(p_\E^2+m^2)^4} \nonumber\\
&&-(4im\,{\cal G}_\E^{\mu\nu\rho\alpha\beta}k_\E^\alpha k_\E^\beta k_\E^\rho+4im\,{\cal G}_\E^{\mu\nu\rho}k_\E^2 k_\E^\rho)\int \frac{d^4p_\E}{(2\pi)^4}\frac{1}{(p_\E^2+m^2)^3}.
\end{eqnarray}

In the above expression, let us first calculate the space integral by using the decomposition (\ref{decom}) as well as the replacement (\ref{rep}) and 
\begin{eqnarray}
\hat p^\mu \hat p^\nu \hat p^\lambda \hat p^\rho &\to & \frac{\hat p^4}{D(D+2)}[(\delta^{\mu\nu}-\delta^{\mu0}\delta^{\nu0})(\delta^{\lambda\rho}-\delta^{\lambda0}\delta^{\rho0}) +(\delta^{\mu\lambda}-\delta^{\mu0}\delta^{\lambda0})(\delta^{\nu\rho}-\delta^{\nu0}\delta^{\rho0}) \nonumber\\
&&+(\delta^{\mu\rho}-\delta^{\mu0}\delta^{\rho0})(\delta^{\lambda\nu}-\delta^{\lambda0}\delta^{\nu0})],
\end{eqnarray}
so that we get
\begin{eqnarray}\label{Pik}
\Pi_k^{\mu\nu} &=&  -\frac im\,{\cal G}^{\mu\nu\rho\alpha\beta}k_\alpha k_\beta k_\rho\,G(m) + \frac i{2m}\,{\cal G}^{\mu\nu\rho00}k^2k_\rho\,H_1(m) \nonumber\\
&&+ \frac i{2m}\,{\cal G}^{\mu\nu\rho00}k_0^2k_\rho\,H_2(m),
\end{eqnarray}
or, in other words, also re-installing the electron charge, we arrive at  
\begin{eqnarray}\label{HD}
{\cal L}_k&=&-\frac {e^2}m G(m)\, {\cal G}^{\mu\nu\rho\alpha\beta} A_\mu \partial_\rho\partial_\alpha\partial_\beta A_\nu + \frac {e^2}{2m}H_1(m)\, {\cal G}^{\mu\nu\rho00} A_\mu \partial_\rho\Box A_\nu \nonumber\\
&&+ \frac {e^2}{2m}H_2(m)\, {\cal G}^{\mu\nu\rho00} A_\mu \partial_\rho\partial_0^2 A_\nu,
\end{eqnarray}
where
\begin{subequations}\label{GH}
\begin{eqnarray}
\label{G} G(m) &=& \mu' \int \frac{dp_0}{2\pi}\frac{m^2}{(p_0^2+m^2)^{3-\frac D2}}, \\
\label{H1} H_1(m) &=& -\mu' \int \frac{dp_0}{2\pi}\left[\frac{(D-5)m^2}{(p_0^2+m^2)^{3-\frac D2}}-\frac{(D-6)m^4}{(p_0^2+m^2)^{4-\frac D2}}\right], \\
\label{H2} H_2(m) &=& -2\mu'\int \frac{dp_0}{2\pi}\left[\frac{(D-3)(D-5)m^2}{(p_0^2+m^2)^{3-\frac D2}}-\frac{2(D-5)(D-6)m^4}{(p_0^2+m^2)^{4-\frac D2}}+\frac{(D-6)(D-8)m^6}{(p_0^2+m^2)^{5-\frac D2}}\right]
\end{eqnarray}
\end{subequations}
with $\mu'=\frac 13 2^{1-D}\pi^{-D/2}(\mu^2)^{\frac32-\frac D2}\Gamma\left(3-\frac D2\right)$. As expected, by calculating the time integral, $G(m)=1/24\pi^2$, while $H_1(m)$ and $H_2(m)$ vanish, which is the result previously obtained in Ref.~\cite{Mar}, at zero temperature.

Now, to employ the Matsubara formalism in the expressions (\ref{GH}), as usually, we take $p_0=(n+1/2)2\pi/\beta$ and change $(1/2\pi)\int dp_0\to 1/\beta \sum_n$, so that $G(m)\to G(\xi)$, and so on. In spite of the resulting sums of (\ref{GH}) are all convergent, we will use the Eq.~(\ref{sum}) to obtain the corresponding expressions which vary with temperature $T=\beta^{-1}$.  The same results can also be obtained through the numerical calculation of these resulting sums, as will see.

Nevertheless, we observe that all the exponents of~(\ref{GH}) are out of range of validity, since we have $\lambda\to3/2$, $\lambda\to5/2$, and $\lambda\to7/2$, when we take $D\to3$ in the summations. As in the previous section, for the sums with $\lambda\to3/2$, we must use the recurrence relation (\ref{rc}) in order to put them in the range of validity, i.e. with $\lambda-1\to1/2$ and $\lambda-2\to-1/2$. Now, for the case of $\lambda\to5/2$, we must insert the expression
\begin{eqnarray}\label{rc2}
f_{\lambda-1}(a,b) &=& -\frac1{2a^2}\frac{2\lambda-5}{\lambda-2}f_{\lambda-2}(a,b) - \frac1{4a^2}\frac1{(\lambda-3)(\lambda-2)}\frac{\partial^2}{\partial b^2}f_{\lambda-3}(a,b)
\end{eqnarray}
into the first term of~(\ref{rc}), so that we then have $\lambda-2\to1/2$ and $\lambda-3\to-1/2$. Finally, for the sums with $\lambda\to7/2$, we first insert 
\begin{eqnarray}\label{rc3}
f_{\lambda-2}(a,b) &=& -\frac1{2a^2}\frac{2\lambda-7}{\lambda-3}f_{\lambda-3}(a,b) - \frac1{4a^2}\frac1{(\lambda-4)(\lambda-3)}\frac{\partial^2}{\partial b^2}f_{\lambda-4}(a,b)
\end{eqnarray}
into the first term of~(\ref{rc2}), and then this Eq.~(\ref{rc2}) and~(\ref{rc3}) into the first and second terms of~(\ref{rc}), respectively, so that we now get $\lambda-3\to1/2$ and $\lambda-4\to-1/2$.

\begin{figure}[h]
\includegraphics[scale=1.0]{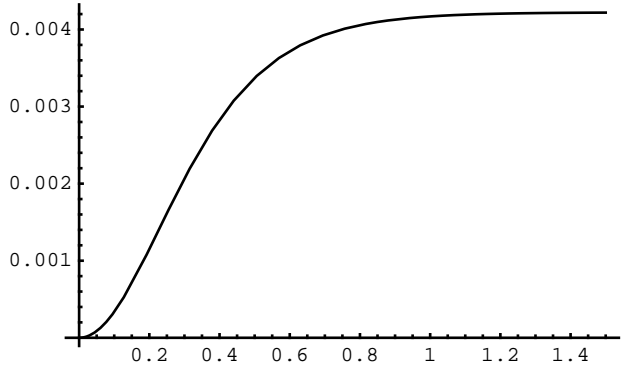}
\caption{Plot of the function $G(\xi)$} \label{fG}
\end{figure}

\begin{figure}[h]
\includegraphics[scale=1.0]{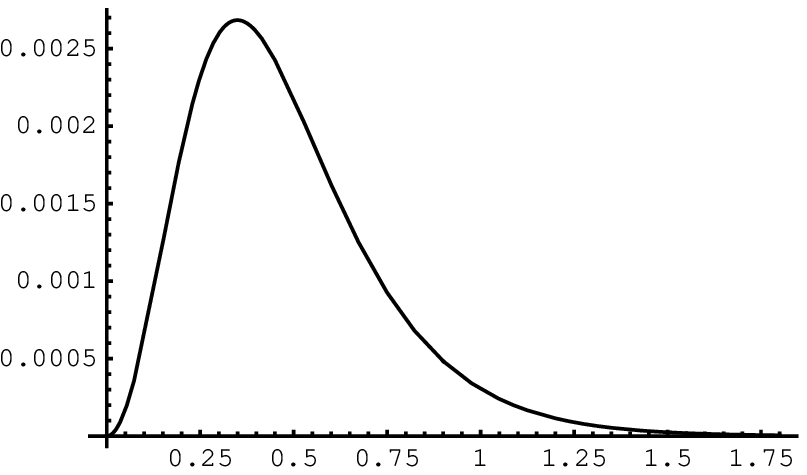}
\caption{Plot of the function $H_1(\xi)$} \label{fH1}
\end{figure}

\begin{figure}[h]
\includegraphics[scale=1.0]{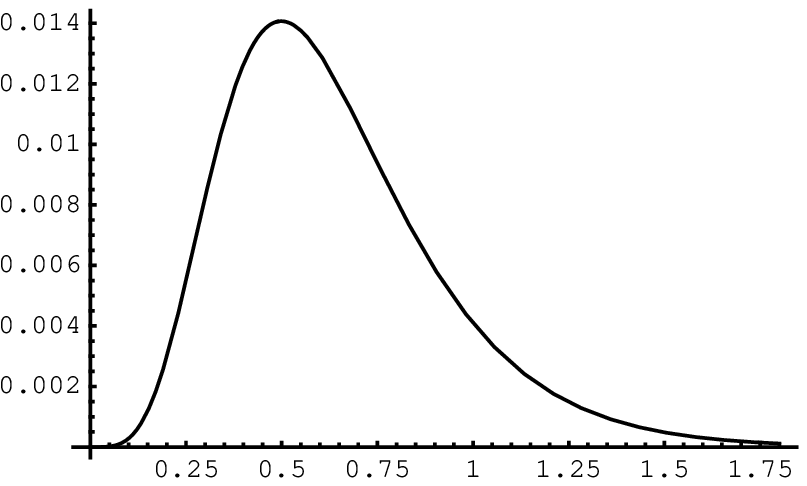}
\caption{Plot of the function $H_2(\xi)$} \label{fH2}
\end{figure}

After all these procedures,  we obtain 
\begin{subequations}\label{GH2}
\begin{eqnarray}
\label{G2}G(\xi) &=& \frac1{24\pi^2} - \frac1{12}\int_{|\xi|}^\infty dz (z^2-\xi^2)^{1/2}\sech^2(\pi\xi)\tanh(\pi\xi), \\
\label{H12}H_1(\xi) &=&  \frac1{12}\int_{|\xi|}^\infty dz (z^2-\xi^2)^{-1/2}\xi^2\,\sech^2(\pi\xi)\tanh(\pi\xi), \\
\label{H22}H_2(\xi) &=&  \frac{\pi^2}{6}\int_{|\xi|}^\infty dz (z^2-\xi^2)^{1/2}\xi^2\,\sech^5(\pi\xi)[\sinh(3\pi\xi)-11\sinh(\pi\xi)].\hspace{1cm}
\end{eqnarray}
\end{subequations}
These functions are depicted in Figs.~\ref{fG}, \ref{fH1}, and \ref{fH2}, respectively. As we can see, in the limit of zero temperature, i.e. $\xi\to\infty$, $G(\xi\to\infty)\to1/24\pi^2$, $H_1(\xi\to\infty)\to0$, and $H_2(\xi\to\infty)\to0$,  as expected. Surprisingly, in the limit of high temperature ($\xi\to0$) all the expressions (\ref{GH2}) vanish. In fact, when we analyze the expressions (\ref{GH}), we note that they are strongly suppressed by the temperature. Another interesting point is that the asymptotic limits of (\ref{G2}) are the opposite of (\ref{F}): at zero temperature only the higher-derivative term is induced, whereas at high temperature only the Chern-Simons term appears. 

As we have mentioned before, the same plot of functions (\ref{GH2}) (presented in Figs.~\ref{fG}, \ref{fH1}, and \ref{fH2}) can also be numerically calculated from the alternative expressions 
\begin{subequations}\label{GH3}
\begin{eqnarray}
\label{G3}G(\xi) &=& \sum_n\frac{\xi^2}{48\pi^2}\left[\left(n+1/2\right)^2+\xi^2\right]^{-3/2},\\
\label{H13}H_1(\xi) &=&  \sum_n\frac{\xi^2}{24\pi^2}\left[\left(n+1/2\right)^2+\xi^2\right]^{-3/2} - \sum_n\frac{\xi^4}{16\pi^2}\left[\left(n+1/2\right)^2+\xi^2\right]^{-5/2},\\
\label{H23}H_2(\xi) &=&  \sum_n\frac{\xi^4}{2\pi^2}\left[\left(n+1/2\right)^2+\xi^2\right]^{-5/2} - \sum_n\frac{5\xi^6}{8\pi^2}\left[\left(n+1/2\right)^2+\xi^2\right]^{-7/2},
\end{eqnarray}
\end{subequations}
which are obtained when we simply take $D=3$ in the expressions (\ref{GH}). The main disadvantage of this approach is in the procedure for obtaining the asymptotic limits, zero and high temperatures, in which, on the other hand, are easily calculated from Eqs.~(\ref{GH2}).

Finally, we would like to comment on the last two terms of Eq.~(\ref{HD}), controlled by the two functions $H_1(\xi)$ and $H_2(\xi)$, respectively, which are responsible for the induction of a noncovariant higher-derivative Chern-Simons term. Analyzing the plots of these functions, observe that $H_1(\xi)$ and $H_2(\xi)$ increase up to a certain temperature, and then decrease as the temperature grows to infinity. This noncovariant higher-derivative term is consistent with the previously induced Chern-Simons term (\ref{CS}), however, it vanishes at zero and high temperatures. 

\section{Summary}\label{su}

In this work we have studied the effect of a thermal bath on the structure of  the higher-derivative term (\ref{k5}) and of the Chern-Simons term (\ref{k3}), generated by integrating out the fermions coupled to the coefficient $g^{\mu\nu\rho}$. With regard to the Chern-Simons term, we have obtained a term that is nonzero only at finite temperature (\ref{CS}), where the corresponding spacelike coefficient can be written as $b_i=\frac12 \epsilon_{ijk0}g^{jk0}$. Therefore, as the coefficients $g^{\mu\nu\rho}$ and $b^\mu$ always appear together, e.g. through a field redefinition, we also expect a radiative contribution coming from $g^{\mu\nu\rho}$ for the Lorentz-violating Chern-Simons term. Now, with regard to the higher-derivative term, we have found a term that is finite at zero temperature, however, it goes to zero as the temperature grows to infinity (see the first term of~(\ref{HD})). As expected, $G(\xi\to\infty)\to1/24\pi^2$, indicating that the perturbative calculation adopted here is consistent with the early result obtained in~\cite{Mar}. Besides these terms, we have also generated a higher-derivative Chern-Simons term, only at finite temperature, nevertheless, it vanishes at high temperature (see the last two terms of~(\ref{HD})). This behavior of the higher-derivative terms, at high temperature, is mainly due to the fact that the summations (\ref{GH3}) are strongly suppressed by the temperature. Thus, as this seems to be an inherent characteristic of higher-derivative terms, we believe that only the operators of mass dimension $d\le4$ survive in the limit of $T\to\infty$.

\vspace{.5cm}
{\bf Acknowledgements.} This work was supported by Conselho Nacional de Desenvolvimento Cient\'{\i}fico e Tecnol\'{o}gico (CNPq).

\end{document}